\documentclass[twocolumn,showpacs,preprintnumbers,amsmath,amssymb]{revtex4}

\usepackage{graphicx}
\usepackage{dcolumn}
\usepackage{bm}

\begin{document}

\title{Equilibrium long-ranged charge correlations at the interface \\
between media coupled to the electromagnetic radiation}
\author{Bernard Jancovici}
\email{Bernard.Jancovici@th.u-psud.fr}
\author{Ladislav \v{S}amaj}
\altaffiliation[On leave from ]
{Institute of Physics, Slovak Academy of Sciences,
Bratislava}
\email{Ladislav.Samaj@savba.sk}
\affiliation{Laboratoire de Physique Th\'eorique,
Universit\'e de Paris-Sud,
91405 Orsay Cedex, France\footnote{Unit\'e Mixte de Recherche No 8627-CNRS}}

\date{\today}

\begin{abstract}
We continue studying long-ranged quantum correlations of surface charge
densities on the interface between two media of distinct dielectric functions 
which are in thermal equilibrium with the radiated electromagnetic field.
Two regimes are considered: the non-retarded one with the speed of light $c$ 
taken to be infinitely large and the retarded one with finite value of $c$.
The analysis is based on results obtained by using fluctuational 
electrodynamics in
[L. \v{S}amaj and B. Jancovici, Phys. Rev. E {\bf 78}, 051119 (2008)].
Using an integration method in the complex plane and the general analytic 
properties of dielectric functions in the frequency upper half-plane,
we derive explicit forms of prefactors to the long-range decay of
the surface charge correlation functions for all possible media
(conductor, dielectric, vacuum) configurations.
The main result is that the time-dependent quantum prefactor in the retarded 
regime takes its static classical form, for any temperature.  
\end{abstract}


\maketitle

\section{Introduction}
In this paper, we continue studying long-ranged quantum correlations 
of surface charge densities on the interface between two distinct media,
initiated in Refs. \cite{SJ,JS}.
The model, formulated in the three-dimensional (3D) Cartesian space 
of coordinates $(x,y,z)$, is inhomogeneous say along the first coordinate 
$x$ (see Fig. 1).
The two semi-infinite media with the frequency-dependent dielectric functions 
$\epsilon_1(\omega)$ and $\epsilon_2(\omega)$ (the magnetic permeability 
$\mu=1$) are localized in the half-spaces $x>0$ and $x<0$, respectively.
The interface is the plane $x=0$, a point on the interface is 
${\bf R}=(0,y,z)$.
The different electric properties of the media give rise to a surface
charge density which must be understood as being the microscopic volume 
charge density integrated on some microscopic depth.
It is related to the discontinuity of the $x$ component of the electric field 
on the interface. 
Denoting by $\sigma(t,{\bf R})$ the surface charge density at time $t$
and at a point ${\bf R}$, the (symmetrized) two-point correlation function,
at times different by $t$, reads
\begin{equation} \label{1}
S(t,{\bf R}) \equiv \frac{1}{2} \langle\sigma(t,{\bf R})
\sigma(0,{\bf 0})+\sigma(0,{\bf 0})\sigma(t,{\bf R})\rangle^{\rm T} , 
\end{equation}
where $\langle\cdots\rangle^{\rm T}$ represents a truncated statistical 
average at the inverse temperature $\beta$ (we exclude the case of
zero temperature, $\beta\to\infty$).
We are interested in the behavior of the correlation function (\ref{1}) 
at distances on the interface $R=\vert {\bf R}\vert$ large compared to 
the microscopic length scales (like the particle correlation function).
The static case of zero time difference $t=0$ between the two distinct points
is simpler than the one with $t\ne 0$ and so the two cases are treated 
separately.   

\begin{figure}
\begin{center}
\includegraphics[width=0.35\textwidth,clip]{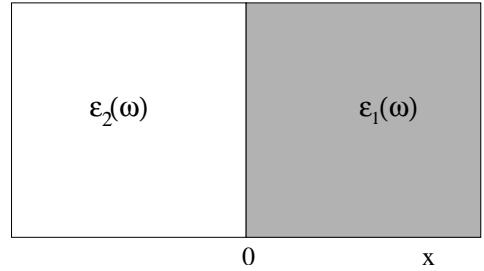}
\caption{Two semi-infinite media characterized by dielectric functions
$\epsilon_1(\omega)$ and $\epsilon_2(\omega)$.}
\end{center}
\end{figure}

The two media configuration studied so far was restricted to a conductor,
localized say in the half-space $x>0$ with the dielectric function
$\epsilon_1(\omega)\equiv \epsilon(\omega)$, in contact with vacuum
of the dielectric constant $\epsilon_2(\omega)=1$ (the vacuum part of 
the space is equivalent to a hard wall impenetrable to charged particles 
forming the conductor).
In some theoretical studies, the dielectric function is approximated 
by a simple one-resonance Drude formula \cite{Jackson}
\begin{equation} \label{2}
\epsilon(\omega) = 1 + \frac{\omega_p^2}{\omega_0^2-\omega(\omega+i\eta)} ,
\end{equation}
where $\omega_p$ is the plasma frequency, $\eta$ the dissipation constant 
and $\omega_0$ the oscillation frequency of harmonically bound charges:
$\omega_0=0$ for conductors and $\omega_0\ne 0$ for dielectrics.
In \cite{SJ} and \cite{JS}, we applied (\ref{2}) to the jellium model 
(sometimes called the one-component plasma), i.e. a system of pointlike 
particles of charge $e$, mass $m$, and bulk number density $n$, immersed in 
a uniform neutralizing background of charge density $-e n$.
The dynamical properties of the jellium have a special feature:
There is no viscous damping of the long-wavelength plasma oscillations
for identically charged particles, so that $\eta\to 0^+$ in (\ref{2}).
The frequencies of non-retarded nondispersive long-wavelength collective 
modes, namely $\omega_p$ of the bulk plasmons and $\omega_s$ of 
the surface plasmons, are given by
\begin{equation} \label{3}
\omega_p = \left( \frac{4\pi n e^2}{m} \right)^{1/2} , \qquad
\omega_s = \frac{\omega_p}{\sqrt{2}} .
\end{equation}
For other conductors and dielectrics, we shall use only general properties
of $\epsilon(\omega)$, without restricting ourselves to the Drude model
(\ref{2}).

The problem of a conductor in contact with vacuum was studied
in the past, with an increasing level of physical complexity 
and their range of validity.
The problem can be treated as classical or quantum, in the non-retarded 
or retarded regime.
In all cases, the asymptotic large-distance behavior of the surface 
charge correlation function (\ref{1}) exhibits a long-ranged tail of type
\begin{equation} \label{4}
\beta S(t,{\bf R}) \sim \frac{h(t)}{R^3} , \qquad R\to\infty ,
\end{equation}
where the form of the prefactor function $h(t)$ depends on the physical 
theory which is used.
It is useful to introduce the Fourier transform
\begin{equation} \label{5}
S(t,{\bf q}) = \int d^2 R \exp(i{\bf q}\cdot{\bf R}) S(t,{\bf R}) ,  
\end{equation}
with ${\bf q}=(q_y,q_z)$ being a 2D wave vector.
Since, in the sense of distributions, the 2D Fourier transform of 
$1/R^3$ is $-2\pi q$, a result equivalent to (\ref{4}) is that 
$\beta S(t,{\bf q})$ has a kink singularity at ${\bf q}={\bf 0}$, 
behaving like
\begin{equation} \label{6}
\beta S(t,q) \sim - 2 \pi h(t) q , \qquad q\to 0 .  
\end{equation}

$\bullet$ {\bf Classical non-retarded regime:}
Let the conductor be modelled by a classical Coulomb fluid composed
of charged particles with the instantaneous Coulomb interactions.
By a microscopic analysis \cite{Jancovici82}, the long-range decay of
the static surface correlation function was found such that
\begin{equation} \label{7}
h_{\rm cl}^{(\rm nr)}(0) = - \frac{1}{8\pi^2} ,
\end{equation} 
where the subscript ``cl'' means ``classical'' and the upperscript ``nr''
means ``non-retarded'' (i.e. considered without relativistic effects
associated with the finiteness of the speed of light $c$).
The same result has been obtained later \cite{Jancovici95} by simple 
macroscopic arguments based on a combination of the linear response theory 
and the electrostatic method of images. 
Note the universal form of $h_{\rm cl}(0)$, independent of the composition
of the Coulomb fluid.

$\bullet$ {\bf Quantum non-retarded regime:}
The extension of the static result (\ref{7}) to a quantum Coulomb fluid,
modelled by the jellium, was accomplished in Ref. \cite{Jancovici85a}.
The absence of damping was crucial in the treatment using long-wavelength
collective modes, the bulk and surface plasmons with frequencies 
given in (\ref{3}).
The Maxwell equations, obeyed by the plasmons, were considered in 
the non-retarded (non-relativistic) regime with the speed of light $c$ 
taken to be infinitely large, $c=\infty$, ignoring in this way magnetic 
forces acting on the charged particles.
The obtained time-dependent result has the nonuniversal form
\cite{Jancovici85a,Jancovici85b,Jancovicicor}
\begin{subequations}
\begin{eqnarray}
h_{\rm qu}^{(\rm nr)}(t) & = & - \frac{1}{8\pi^2} \nonumber \\ & & \times 
\left[ 2 g(\omega_s)\cos(\omega_s t) - g(\omega_p)\cos(\omega_p t) \right] , 
\label{8a} \\
\quad g(\omega) & = & \frac{\beta\hbar\omega}{2} 
\coth\left( \frac{\beta\hbar\omega}{2} \right) , \label{8b}
\end{eqnarray}
\end{subequations}
where the subscript ``qu'' means ``quantum''.
According to the correspondence principle, a quantum system admits
the classical statistical description in the high-temperature limit 
$\beta\hbar\to 0$.
In this limit, the function $g(\omega)=1$ for any $\omega$ and
the quantum formula (\ref{8a}) reduces to the classical non-retarded one
\begin{equation} \label{9}
h_{\rm cl}^{(\rm nr)}(t) = - \frac{1}{8\pi^2} 
\left[ 2 \cos(\omega_s t) - \cos(\omega_p t) \right] . 
\end{equation}
For $t=0$, we recover the classical static formula (\ref{7}).

$\bullet$ {\bf Quantum retarded regime:}
In the previous paper \cite{SJ}, we studied the surface charge correlations
taking into account retardation ($c$ is assumed finite) and the quantum 
nature of both the jellium and the radiated electromagnetic (EM) field
which are in thermal equilibrium. 
In other words, the quantum particles are fully coupled to both electric 
and magnetic parts of the radiated EM field.
By using Rytov's fluctuational electrodynamics \cite{Rytov,LL} we showed
that there are two regions of distances $R$ on the interface:
the intermediate one given by the inequalities 
$\lambda_{\rm ph}\ll R\ll c/\omega_p$ ($\lambda_{\rm ph}\propto \beta\hbar c$ 
stands for the thermal de Broglie wavelength of photon), 
where the non-retarded result (\ref{8a}), (\ref{8b}) 
applies, and the strictly asymptotic one given by the inequality 
$c/\omega_p\ll R$, where a retarded result applies. 
After long calculations in \cite{SJ}, a bit shortened through 
the alternative method of \cite{JS}, a very simple form of the retarded 
result was found
\begin{equation} \label{10}
h_{\rm qu}^{(\rm r)}(t) = - \frac{1}{8\pi^2} .
\end{equation}
Here, the upperscript ``r'' means ``retarded''.
We see that, for any temperature $\beta$ and time $t$, the inclusion of 
retardation effects causes the prefactor function $h(t)$ to take its 
universal static classical form (\ref{7}), independent of $\hbar$ and $c$. 
The formula (\ref{10}) does not change in the classical limit 
$\beta\hbar\to 0$, i.e.
\begin{equation} \label{11}
h_{\rm cl}^{(\rm r)}(t) = - \frac{1}{8\pi^2} .
\end{equation}
The presence or the absence of magnetic fields is important in two-point 
classical statistical averages, taken at two different times
\cite{Alastuey08}.  
Therefore, it is not surprising that the classical retarded formula (\ref{11}) 
and the classical non-retarded one (\ref{9}) do not coincide with one another
for nonzero time differences, 
$h_{\rm cl}^{(\rm r)}(t) \ne h_{\rm cl}^{(\rm nr)}(t)$ for $t\ne 0$. 
On the other hand, for $t=0$ we have 
\begin{equation} \label{12}
h_{\rm cl}^{(\rm r)}(0) = h_{\rm cl}^{(\rm nr)}(0) .
\end{equation}
This equality is in agreement with the Bohr-van Leeuwen theorem
\cite{Bohr,Leeuwen} about an effective elimination of magnetic degrees
of freedom from statistical averages (with zero time differences among 
the fixed points in the coordinate part of the configuration space) 
of classical systems; for a detailed treatment of this subject,
see Ref. \cite{Alastuey00}. 

In the previous papers \cite{SJ,JS}, we used the fact, special to the jellium
model, that there is no damping for small wave numbers. 
The question whether the crucial formula (\ref{10}) is still valid for 
a conductor with dissipation was left as an open problem. 
There are still many other unsolved physical situations which deserve 
attention.
What happens in the case of a general dielectric in contact with vacuum?
Filling the vacuum region by a material medium, other types of contacts 
are possible, like conductor - conductor, dielectric - dielectric, 
conductor - dielectric.
Another kind of problem is the algebraic complicacy connected with 
the derivation of the result (\ref{10}) for the jellium with the relatively
simple form of the dielectric constant.
Does there exists a simple method for evaluating the large-distance
asymptotics of the surface charge density correlation function which 
is applicable to the jellium as well as to other more complicated systems? 
All asked questions are answered in the present paper.

The generalization of the formalism to contacts between all kinds of 
materials, defined by their dielectric functions, has already 
been done in Ref. \cite{SJ}.
The true problem is the mathematical handling of the final (non-retarded
or retarded) formula for the Fourier transform $\beta S(t,q)$, 
written as an integral over real frequencies, to deduce the small-$q$ behavior
(\ref{6}).
Here, we accomplish the task first by extending the integration over real 
frequencies to a contour integration in the complex frequency plane
and then using integration techniques in the complex plane together with
the known analytic properties of dielectric functions in the frequency upper
half-plane.
In short, the time-dependent retarded results maintain the simplicity
of the jellium formula (\ref{10}) and involve only dielectric functions 
of media in contact at zero frequency.
The non-retarded results are complicated and available only, in general, 
as infinite series over Matsubara frequencies. 

The article is organized as follows. 
Sec. II is a generalization of the classical static result (\ref{7}) 
for conductor to a dielectric.
The result will serve us as a check of more general calculations.
Sec. III summarizes briefly general analytic properties of dielectric 
functions in the complex frequency upper half-plane which are necessary for
the derivation of our basic results.
Sec. IV applies the new method for the calculation of $\beta S(0,{\bf q})$.
Sec. V does the same for $\beta S(t,{\bf q})$. 
Sec. VI is a Conclusion.

\section{Static correlations for classical dielectrics}
It has been known for a long time \cite{Jancovici82} that the classical 
surface charge correlations on a conductor, made of particles interacting 
through the Coulomb law and bounded by a plane wall, are long-ranged. 
We prefer to use the macroscopic language \cite{Jancovici95}. 
If ${\bf R}$ and ${\bf R}'$ are two points on the wall, the classical 
correlation, for distances $\vert {\bf R}-{\bf R}'\vert$ large compared to 
the microscopic scale, behaves like 
\begin{equation} \label{13}
\beta S({\bf R}-{\bf R}') \equiv
\beta\langle\sigma({\bf R})\sigma({\bf R}')\rangle \sim
-\frac{1}{8\pi^2{\vert \bf R}-{\bf R}'\vert^3}  
\end{equation}
(we assume that the conductor is uncharged, $\langle\sigma({\bf R})\rangle=0$).

A generalization of (\ref{13}) for the case of a dielectric of static
dielectric constant $\epsilon(0)=\epsilon_0$ bounded by a plane can be 
found by the same method which has been used in \cite{Jancovici95}. 
$\sigma({\bf R})$ is related to the discontinuity of the $x$ component of the
electric field on the wall. 
We call $E_x^{\rm out(in)}({\bf R})$ the limit of that field component as 
${\bf R}$ is 0 from the outside (inside) of the dielectric. 
The correlation of the surface charge densities is   
\begin{eqnarray}
\langle\sigma({\bf R})\sigma({\bf R}')\rangle & = & \frac{1}{(4\pi)^2}
\langle [E_x^{\rm in}({\bf R})-E_x^{\rm out}({\bf R})] \nonumber  \\ & & 
\times [E_x^{\rm in}({\bf R}')-E_x^{\rm out}({\bf R}')] \rangle .\label{14}
\end{eqnarray}

If a test infinitesimal charge is introduced at ${\bf r}=(x,0,0)$, $x>0$ 
in the dielectric, the electric potential \emph{created by the dielectric} 
at a point ${\bf r}'=(x',y',z')$, $x'>0$ in the dielectric is given by 
the method of images as \cite{Jackson} 
\begin{equation} \label{15}
\langle\phi({\bf r}')\rangle_q = \frac{1}{\epsilon_0}
\left[ \frac{q}{\vert {\bf r}'-{\bf r}\vert} -
\frac{q(1-\epsilon_0)}{(1+\epsilon_0){\vert \bf r}'-{\bf r}^*\vert}\right] 
-\frac{q}{\vert {\bf r}'-{\bf r}\vert },  
\end{equation}
where ${\bf r}^*=(-x,0,0)$ is the image of ${\bf r}$. 
The last term in (\ref{15}) is the potential created by $q$, 
which should not be included in the potential created by the dielectric. 
The linear response theory relates the response (\ref{15}) to 
the unperturbed correlation function between the additional Hamiltonian 
$q\phi({\bf r})$ and $\phi({\bf r}')$, giving
\begin{equation} \label{16}
\langle\phi({\bf r}')\rangle_q  =
- \beta q\langle\phi({\bf r})\phi({\bf r}')\rangle .  
\end{equation}
Since the $x$ component of the electric field is $E_x({\bf r})=
-(\partial/\partial x)\phi({\bf r})$, using (\ref{15}) and (\ref{16}) we can
obtain the correlation of the $x$ component of the electrical fields inside
the dielectric at the wall
\begin{equation} \label{17}
\beta\langle E_x^{\rm in}({\bf R})E_x^{\rm in}({\bf R}')\rangle = 
\left(\frac{2}{1+\epsilon_0}-\frac{2}{\epsilon_0}+1\right)
\frac{1}{\vert {\bf R}-{\bf R}' \vert^3}.  
\end{equation}

Similar calculations give the correlation outside the dielectric
\begin{equation} \label{18}
\beta\langle E_x^{\rm out}({\bf R})E_x^{\rm out}({\bf R}')\rangle =
\frac{1-\epsilon_0}{1+\epsilon_0}\frac{1}{\vert
{\bf R}-{\bf R}' \vert^3}
\end{equation}
and the cross-correlation  
\begin{equation} \label{19}
\beta\langle E_x^{\rm in}({\bf R})E_x^{\rm out}({\bf R}')\rangle =
-\frac{1-\epsilon_0}{1+\epsilon_0}\frac{1}{\vert {\bf R}-{\bf R}' \vert^3} .
\end{equation}
Using (\ref{17}), (\ref{18}) and (\ref{19}) in (\ref{14}), we obtain
\begin{equation} \label{20}
\beta\langle\sigma({\bf R})\sigma({\bf R}')\rangle \sim
-\frac{1}{8\pi^2}\left( \frac{1}{\epsilon_0}+1-\frac{4}{1+\epsilon_0} \right)
\frac{1}{\vert {\bf R}-{\bf R}' \vert^3},     
\end{equation}
which is the wanted generalization for a dielectric. 
With regard to the definition (\ref{4}), we have the nonuniversal result
\begin{equation} \label{21}
h_{\rm cl}(0) = - \frac{1}{8\pi^2}
\left( \frac{1}{\epsilon_0}+1-\frac{4}{1+\epsilon_0} \right) .
\end{equation}
The value of $\epsilon(0)$ is infinite for any kind of conductor and 
we retrieve (\ref{13}) from (\ref{20}), or (\ref{7}) from (\ref{21}).
This fact explains the universality of the static $h_{\rm cl}$ for 
conductors. 

\section{Analytic properties of dielectric functions}
Analytic properties of dielectric functions are described in many textbooks
\cite{Jackson,LL,LP}.
They apply to an arbitrary dielectric function of real materials,
including the idealized Drude formula (\ref{2}). 
We shall mention only those properties which are important in the
derivation of our basic results; the proofs of theorems are given in 
the above textbooks.
The vacuum case $\epsilon(\omega)=1$ is excluded from the discussion.

Due to the causal relation between the displacement ${\bf D}$ and
the electric field ${\bf E}$, the dielectric function of every medium
can be expressed as
\begin{equation} \label{22}
\epsilon(\omega) = 1 + \int_0^{\infty} d\tau e^{i\omega\tau} G(\tau) ,
\end{equation} 
where the function $G(\tau)$ is finite for all values of $\tau$,
including zero.
In particular, $G(\tau)$ tends to 0 as $\tau\to\infty$ for dielectrics
and it tends to $4\pi\sigma$ ($\sigma$ is the conductivity)
as $\tau\to\infty$ for conductors.
The relation (\ref{22}) has several important consequences.

Let us first consider the frequency $\omega$ to be purely real.
It follows from (\ref{22}) that $\epsilon^*(\omega) = \epsilon(-\omega)$.
Denoting $\epsilon(\omega) = \epsilon'(\omega) + i \epsilon''(\omega)$,
where both the real $\epsilon'(\omega)$ and imaginary $\epsilon''(\omega)$
parts are real numbers, we thus have
\begin{equation} \label{23}
\epsilon'(\omega) = \epsilon'(-\omega) , \qquad
\epsilon''(\omega) = - \epsilon''(-\omega) .
\end{equation}
For any real material medium with absorption it holds
\begin{equation} \label{24}
\epsilon''(\omega) > 0 (<0)  \quad {\rm for} \quad
\omega > 0 (<0) ;
\end{equation}
the sign of $\epsilon'(\omega)$ is not subject to any physical restriction.

If $\omega$ is complex, $\omega = \omega' + i \omega''$,
the representation (\ref{22}) tells us that $\epsilon(\omega)$ is
an analytic function of $\omega$ in the upper half-plane $\omega''>0$.
Apart from a possible pole at $\omega=0$ (for conductors), the analyticity
extends also to the real $\omega$ axis. 

The function $\epsilon(\omega)$ does not take real values at any finite
point in the upper half-plane, except on the imaginary axis.
We can deduce from (\ref{22}) that for any complex $\omega$ it holds
$\epsilon^*(\omega) = \epsilon(-\omega^*)$.
For purely imaginary $\omega=i\omega''$, indeed we find
\begin{equation} \label{25}
\epsilon(i\omega'') = \epsilon^*(i\omega'') \quad \Longrightarrow \quad
\mbox{${\rm Im}\, \epsilon(\omega) = 0$ for $\omega=i\omega''$.}
\end{equation}
Moreover, on the imaginary axis, $\epsilon(\omega)$ decreases monotonically
from $\epsilon_0>1$ (for dielectrics) or from $\infty$ (for conductors)
at $\omega=i0$ to 1 at $\omega=i\infty$.
Hence, in particular, $\epsilon(\omega)$ has no zeros in the upper half-plane.

With regard to the symmetries (\ref{23}) for a real $\omega$, the expansion 
of the dielectric function around the origin $\omega=0$ reads: 
for a conductor with conductivity $\sigma>0$
\begin{equation} \label{26}
\epsilon(\omega) = \frac{4\pi\sigma i}{\omega} + c + O(\omega) ,
\end{equation}
where the sign of the constant $c$ is not restricted;
for a dielectric medium of static dielectric constant $\epsilon_0$
\begin{equation} \label{27}
\epsilon(\omega) = \epsilon_0 + i c \omega + O(\omega^2) , \qquad c>0 ,
\end{equation}
where the positive sign of the constant $c$ is fixed by 
the physical requirements (\ref{24}).

In the limit $\vert\omega\vert\to \infty$, a Taylor series expansion
of $G(\tau)$ around $\tau=0^+$ in Eq. (\ref{22}) implies that, 
for both conductors and dielectrics,
\begin{subequations}
\begin{eqnarray}
{\rm Re}\left[ \epsilon(\omega)-1\right] & = & 
O\left( \frac{1}{\omega^2} \right) , \label{28a} \\
{\rm Im}\, \epsilon(\omega) & = & 
O\left( \frac{1}{\omega^3} \right) . \label{28b}
\end{eqnarray}
\end{subequations}

\section{Static charge correlations}
In the general case of the plane contact between two media of dielectric 
functions $\epsilon_1(\omega)$ and $\epsilon_2(\omega)$ pictured in Fig.1,
the quantum formula for the Fourier transform of the surface charge correlation
function (\ref{5}) in the long-wavelength limit $q\to 0$ was derived in 
Ref. \cite{SJ}.
Its static $t=0$ version can be reexpressed as
\begin{equation} \label{29}
\beta S_{\rm qu}(0,q) = \int_{-\infty}^{\infty} \frac{d\omega}{\omega}
{\rm Im}\, f(\omega) ,
\end{equation}
where the form of the function $f(\omega)$ depends on the considered
(retarded or non-retarded) regime.
In the retarded case, $f(\omega)\equiv f_{\rm qu}^{(\rm r)}(\omega)$
is given by two equivalent representations
\begin{eqnarray}
f_{\rm qu}^{(\rm r)}(\omega) & = & \frac{q^2}{4\pi^2} g(\omega)
\frac{1}{\kappa_1(\omega) \epsilon_2(\omega) + 
\kappa_2(\omega) \epsilon_1(\omega)} \nonumber \\ & & \times
\frac{[\epsilon_1(\omega)-\epsilon_2(\omega)]^2}{\epsilon_1(\omega)
\epsilon_2(\omega)} \nonumber \\ 
& = & \frac{q^2}{4\pi^2} g(\omega)
\frac{\kappa_1(\omega)\epsilon_2(\omega) 
- \kappa_2(\omega)\epsilon_1(\omega)}{q^2
[\epsilon_1(\omega)+\epsilon_2(\omega)]
-\omega^2\epsilon_1(\omega)\epsilon_2(\omega)/c^2} \nonumber \\ & & \times
\left( \frac{1}{\epsilon_1(\omega)} - \frac{1}{\epsilon_2(\omega)} \right) ,
\label{30}
\end{eqnarray}
where $g(\omega)$ is defined in (\ref{8b}) and the (complex) inverse lengths
$\kappa_1$ and $\kappa_2$, one for each of the half-space regions, are given by
\begin{equation} \label{31}
\kappa_{1,2}^2(\omega) = q^2 - \frac{\omega^2}{c^2} \epsilon_{1,2}(\omega) ,
\qquad {\rm Re}\, \kappa_{1,2}(\omega) > 0 .
\end{equation}
In the non-retarded case, $f(\omega)\equiv f_{\rm qu}^{(\rm nr)}(\omega)$
is obtained from (\ref{30}) by setting the speed of light $c\to\infty$.
Since $\kappa_1=\kappa_2=q$ in this limit, we get
\begin{equation} \label{32}
f_{\rm qu}^{(\rm nr)}(\omega) = \frac{q}{4\pi^2} g(\omega)
\left( \frac{1}{\epsilon_1(\omega)} + \frac{1}{\epsilon_2(\omega)} 
- \frac{4}{\epsilon_1(\omega)+\epsilon_2(\omega)} \right) .
\end{equation}
The derivation procedures outlined below are applicable to both retarded
and non-retarded regimes and so, whenever possible, we use the simplified 
notation $f(\omega)$ to cover both functions $f_{\rm qu}^{(\rm r)}(\omega)$ 
and $f_{\rm qu}^{(\rm nr)}(\omega)$.

For a real frequency $\omega$, the symmetry relation 
$\epsilon^*(\omega) = \epsilon(-\omega)$ implies that
$\kappa^*(\omega) = \kappa(-\omega)$.
Consequently, $f^*(\omega) = f(-\omega)$, i.e.
\begin{equation} \label{33}
{\rm Re}\, f(\omega) = {\rm Re}\, f(-\omega) , \qquad
{\rm Im}\, f(\omega) = - {\rm Im}\, f(-\omega) .
\end{equation}
As $\omega\to 0$,
\begin{subequations}
\begin{eqnarray}
{\rm Im}\, f(0) & = & 0 , \label{34a} \\
{\rm Re}\, f(0) & = & \frac{q}{4\pi^2} 
\left( \frac{1}{\epsilon_1(0)} + \frac{1}{\epsilon_2(0)} 
- \frac{4}{\epsilon_1(0)+\epsilon_2(0)} \right) .\phantom{aaaa} \label{34b}
\end{eqnarray}
\end{subequations}
If $\omega$ is complex and $\vert\omega\vert \to\infty$, with the aid
of the asymptotic relations (\ref{28a}) and (\ref{28b}) we find that
\begin{equation} \label{35}
\lim_{\vert\omega\vert\to\infty} f(\omega) = 0 .
\end{equation}

\begin{figure}
\begin{center}
\includegraphics[width=0.45\textwidth,clip]{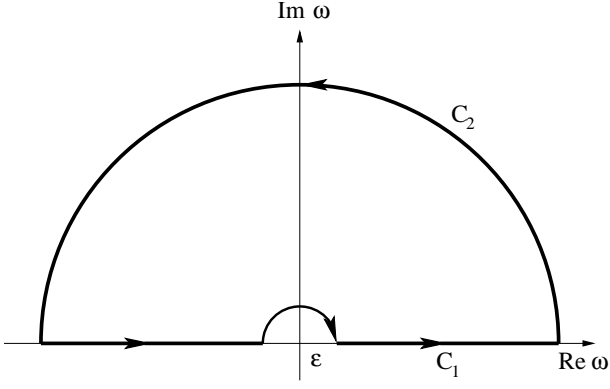}
\caption{The contour in the complex frequency plane for $t=0$.}
\end{center}
\end{figure}

It is complicated to calculate the correlation function directly from
the formula (\ref{29}) by expressing the imaginary part of $f(\omega)$, 
then integrating over $\omega$ and finally taking the $q\to 0$ limit.
We shall find the value of the integral of interest in another way, 
by using integration techniques in the complex plane 
and the analytic properties of dielectric functions, summarized in 
the previous section.

We can change slightly the path of integration, writing
\begin{equation} \label{36}
\int_{C_1} \frac{d\omega}{\omega} f(\omega) = - i \pi f(0)
+ {\cal P} \int_{-\infty}^{\infty} \frac{d\omega}{\omega} f(\omega) ,
\end{equation}
where ${\rm C_1}$ is the path following the real axis, except it goes 
around the origin $\omega=0$ in a small semicircle in complex upper half-plane
whose radius $\varepsilon$ tends to zero (Fig.2).
The first term on the rhs of (\ref{36}) is the contribution of 
the negatively oriented semicircle around the origin, ${\cal P}$ denotes 
the Cauchy principal value avoiding the origin,
\begin{equation} \label{37}
{\cal P} \int_{-\infty}^{\infty} \frac{d\omega}{\omega} f(\omega) \equiv
\lim_{\varepsilon\to 0} \left[
\int_{-\infty}^{-\varepsilon} \frac{d\omega}{\omega} f(\omega) +
\int_{\varepsilon}^{\infty} \frac{d\omega}{\omega} f(\omega) \right] .
\end{equation} 
It is easy to see from the symmetry relations (\ref{33}) and the equality
(\ref{34a}) that
\begin{equation} \label{38}
{\cal P} \int_{-\infty}^{\infty} \frac{d\omega}{\omega} f(\omega) = i
\int_{-\infty}^{\infty} \frac{d\omega}{\omega} {\rm Im}\, f(\omega) 
\end{equation} 
We can close the path $C_1$ by a semicircle at infinity $C_2$ 
(see Fig. 2) along which the integral 
$\int_{\rm C_2} d\omega\, f(\omega)/\omega$ is 0 because of 
the asymptotic relation (\ref{35}). 
Denoting the closed contour as $C$, $C=C_1\cup C_2$, and applying the
operation Im to both sides of the relation (\ref{36}), we arrive at
\begin{equation} \label{39}
\int_{-\infty}^{\infty} \frac{d\omega}{\omega} {\rm Im}\, f(\omega) 
= \pi f(0) + {\rm Im}\, \oint_{C} \frac{d\omega}{\omega} f(\omega) . 
\end{equation}
The integral over the contour $C$ can be evaluated by using
the residue theorem at poles $\{ \omega_j \}$ of the function $f(\omega)$ 
in the $\omega$ upper half-plane bounded by $C$,
\begin{equation} \label{40}
{\rm Im}\, \oint_{C} \frac{d\omega}{\omega} f(\omega) =
2\pi \sum_j \frac{{\rm Res}(f,\omega_j)}{\omega_j} ,
\end{equation}
provided that ${\rm Res}(f,\omega_j)/\omega_j$ is real
(which will be the case); ${\rm Res}$ denotes the residue.
The static correlation function (\ref{29}) is expressible as
\begin{equation} \label{41} 
\beta S_{\rm qu}(0,q) =  \pi f(0) +
2\pi \sum_j \frac{{\rm Res}(f,\omega_j)}{\omega_j} ,
\end{equation}
where the value of $f(0)$ is real, given by (\ref{34b}).
The original algebraic task thus reduces to the problem of searching 
for all poles of the function $f(\omega)$ in the $\omega$ upper half-plane. 

Both the retarded (\ref{30}) and non-retarded (\ref{32}) versions
of the $f$-function contain $g(\omega)$ defined in (\ref{8b}).
Since $g(\omega)$ can be expanded in $\omega$ as follows \cite{Gradshteyn}
\begin{equation} \label{42}  
g(\omega) = 1 + \sum_{j=1}^{\infty} \frac{2\omega^2}{\omega^2+\xi_j^2} ,
\qquad \xi_j = \frac{2\pi}{\beta\hbar} j ,
\end{equation}
it has in the upper half-plane an infinite sequence of simple poles 
at the imaginary Matsubara frequencies
\begin{equation} \label{43}
\omega_j = i \xi_j , \qquad {\rm Res}(g,\omega_j) = \omega_j 
\quad (j=1,2,\ldots) .
\end{equation}

The non-retarded $f$-function (\ref{32}) has no further poles in 
the upper half-plane since the dielectric functions $\epsilon_1(\omega)$ 
and $\epsilon_2(\omega)$ do not take there real values at any finite point. 
The retarded $f$-function (\ref{30}) might have some further poles
at points $\omega$ satisfying the equation
\begin{equation} \label{44}
\omega^2 = (cq)^2 \left( \frac{1}{\epsilon_1(\omega)}
+ \frac{1}{\epsilon_2(\omega)} \right) .
\end{equation}
Interestingly, in the case of the jellium in vacuum (with real dielectric
functions), this is just the dispersion relation for the surface plasmons 
(polaritons); for a recent review, see \cite{Pitarke}.
We are interested in the long-wavelength limit $q\to 0$.
If $q=0$, the only solution of Eq. (\ref{44}) is $\omega=0$;
this point is not inside the contour $C$ due to the presence of 
the semicircle around $\omega=0$.
Now, let us study how the solutions of Eq. (\ref{44}) ``glue off'' from
$\omega=0$ when $q$ is infinitesimal, but not identically equal to zero. 
We need the small-$\omega$ expansion of $1/\epsilon(\omega)$.
It follows from the expansions (\ref{26}) and (\ref{27}) that for both
conductors and dielectrics we can write
\begin{equation} \label{45}
\frac{1}{\epsilon(\omega)} = \frac{1}{\epsilon(0)} - i b \omega
+ O(\omega^2) , \qquad b\ge 0 ,
\end{equation}
where the material constant $b$, $b=1/(4\pi\sigma)$ for conductors and
$b=c/\epsilon_0^2$ for dielectrics, is always positive, except for vacuum 
when $b=0$.
For small $q$, Eq. (\ref{44}) thus exhibits two solutions  
\begin{equation} \label{46}
\omega_{\pm} \sim \pm c q \left( \frac{1}{\epsilon_1(0)} + 
\frac{1}{\epsilon_2(0)} \right)^{1/2} - i \frac{(cq)^2}{2} (b_1+b_2) .
\end{equation}
Since $b_1+b_2>0$, the two poles $\omega_{\pm}$ move, as $q$ increases from 0 
to a small positive number, from $\omega=0$ to the lower $\omega$ half-plane, 
i.e. outside of the region enclosed by the $C$-contour.
We conclude that, in both retarded and non-retarded regimes, only 
the poles on the imaginary axis at the Matsubara frequencies (\ref{43}) 
contribute to the static correlation function (\ref{41}), which thus
becomes expressible as follows
\begin{subequations}
\begin{eqnarray} 
\beta S_{\rm qu}(0,q) & = & \frac{q}{4\pi} 
\left( \frac{1}{\epsilon_1(0)} + \frac{1}{\epsilon_2(0)} 
- \frac{4}{\epsilon_1(0)+\epsilon_2(0)} \right) \nonumber \\ & &
+ F(0,q) , \label{47a} \\ 
F(0,q) & = & 2\pi \sum_{j=1}^{\infty} \frac{{\rm Res}(f,i\xi_j)}{i\xi_j} 
\label{47b} .
\end{eqnarray}
\end{subequations}
The first term on the rhs of Eq. (\ref{47a}) is independent of 
$\beta\hbar$ and $c$, the explicit form of the (static) function 
$F(0,q)$ depends on the considered (retarded or non-retarded) regime.

\subsection{Retarded regime}
In the retarded case (\ref{30}), we have
\begin{eqnarray}
F_{\rm qu}^{(\rm r)}(0,q) & = &  \frac{q^2}{2\pi} \sum_{j=1}^{\infty}
\frac{1}{\kappa_1(i\xi_j) \epsilon_2(i\xi_j) +
\kappa_2(i\xi_j) \epsilon_1(i\xi_j)} \nonumber \\ & & \times
\frac{[\epsilon_1(i\xi_j)-\epsilon_2(i\xi_j)]^2}{\epsilon_1(i\xi_j)
\epsilon_2(i\xi_j)} . \label{48}
\end{eqnarray}
We recall from Sec. III that the values of the dielectric functions 
$\epsilon_{1,2}(i\xi_j)$, and consequently of the inverse lengths 
$\kappa_{1,2}(i\xi_j)$ (\ref{31}), are real.
We are interested in the limit $q\to 0$ for which
$\kappa_{1,2}(i\xi_j) = \xi_j \epsilon_{1,2}^{1/2}(i\xi_j)$.
Since $\xi_j\propto j$ and, according to (\ref{28a}), 
$\epsilon(i\xi_j)-1 = O(1/j^2)$, the sum in (\ref{48}) converges.
This means that the function $F_{\rm qu}^{(\rm r)}(0,q)$, being of 
the order $O(q^2)$, becomes negligible in comparison with the first 
term in (\ref{47a}) when $q\to 0$.
In view of the representation (\ref{6}), we find the static prefactor 
associated with the asymptotic decay to be
\begin{equation} \label{49}
h_{\rm qu}^{(\rm r)}(0) = - \frac{1}{8\pi^2} 
\left( \frac{1}{\epsilon_1(0)} + \frac{1}{\epsilon_2(0)} 
- \frac{4}{\epsilon_1(0)+\epsilon_2(0)} \right) .
\end{equation}

Since this expression does not depend on the temperature and $\hbar$,
its classical $\beta\hbar\to 0$ limit is the same, i.e.
\begin{equation} \label{50}
h_{\rm cl}^{(\rm r)}(0) = - \frac{1}{8\pi^2} 
\left( \frac{1}{\epsilon_1(0)} + \frac{1}{\epsilon_2(0)} 
- \frac{4}{\epsilon_1(0)+\epsilon_2(0)} \right) .
\end{equation}

\subsection{Non-retarded regime}
In the non-retarded case (\ref{32}), we have
\begin{eqnarray}
F_{\rm qu}^{(\rm nr)}(0,q) & = & \frac{q}{2\pi} \sum_{j=1}^{\infty}
\left( \frac{1}{\epsilon_1(i\xi_j)} + \frac{1}{\epsilon_2(i\xi_j)}
\right. \nonumber \\ & & \left. 
- \frac{4}{\epsilon_1(i\xi_j)+\epsilon_2(i\xi_j)} \right) . \label{51}
\end{eqnarray}
It is evident from the asymptotic behavior $\epsilon(i\xi_j)-1 = O(1/j^2)$ 
that the sum in (\ref{51}) converges.
The function $F_{\rm qu}^{(\rm nr)}(0,q)$ is of the order $O(q)$ and
its contribution to $\beta S_{\rm qu}^{(\rm nr)}(0,q)$ in (\ref{47a})
is nonzero in the limit $q\to 0$.

The explicit evaluation of the infinite sum over the Matsubara frequencies 
in (\ref{51}) is, in general, very complicated.
As a check of the presented formalism, we reconsider the previously 
studied case of the jellium in contact with vacuum, i.e.
\begin{equation} \label{52}
\epsilon_1(\omega) = 1 - \frac{\omega_p^2}{\omega^2} , \qquad
\epsilon_2(\omega) = 1 .
\end{equation}
Inserting these dielectric functions into (\ref{51}) and 
using the analog of the summation formula (\ref{42}),
\begin{equation} \label{53}
\sum_{j=1}^{\infty} \frac{(\alpha/\pi)^2}{j^2+(\alpha/\pi)^2} =
\frac{1}{2} \left( \alpha \coth\alpha - 1 \right)
\end{equation}
for $\alpha=\beta\hbar\omega_p/2$ and $\beta\hbar\omega_s/2$
$(\omega_s=\omega_p/\sqrt{2})$, we obtain
\begin{equation} \label{54}
F_{\rm qu}^{(\rm nr)}(0,q) = \frac{q}{4\pi}
\left[ 2 g(\omega_s) - g(\omega_p) -1 \right] 
\end{equation}
with $g(\omega)$ defined in (\ref{8b}).
Adding to this function the first term $q/(4\pi)$ in (\ref{47a}),
we reproduce the $t=0$ case of the previous result (\ref{8a}).

In the classical (high-temperature) limit $\beta\hbar\to 0$, each of 
the frequencies $\{ \xi_j \}_{j=1}^{\infty}$ tends to infinity,
the corresponding terms in the summation over $j$ in Eq. (\ref{51}) 
vanish and so $F_{\rm qu}^{(\rm nr)}(0,q) \to 0$.
We are left with only the contribution identical to the retarded
classical result (\ref{50}), i.e.
$h_{\rm cl}^{(\rm nr)}(0) = h_{\rm cl}^{(\rm  r)}(0)$, as it should be.
For the configuration of a conductor, $\epsilon_1(0)=\infty$, in contact
with vacuum, $\epsilon_2(0)=1$, we recover the universal result (\ref{7}).
For the configuration of a dielectric, $\epsilon_1(0)=\epsilon_0$, in contact
with vacuum, $\epsilon_2(0)=1$, we recover our previous result (\ref{21}).
The classical static prefactor $h_{\rm cl}(0)$ to the $1/R^3$ asymptotic decay 
is nonzero for an arbitrary configuration of two distinct media, except for 
the special case of two conductors.
In that special case, the asymptotic decay is of a short-ranged type.

\section{Time-dependent charge correlations}
For an arbitrary time difference $t\ge 0$ between two points on the interface, 
the quantum formula for the Fourier transform of the surface charge 
correlation function (\ref{5}) reads \cite{SJ}
\begin{subequations}
\begin{eqnarray}
\beta S_{\rm qu}(t,q) & = & \int_{-\infty}^{\infty} \frac{d\omega}{\omega}
e^{-i\omega t} {\rm Im}\, f(\omega) , \label{55a} \\
& = & \int_{-\infty}^{\infty} \frac{d\omega}{\omega}
\cos(\omega t) {\rm Im}\, f(\omega) , \label{55b}
\end{eqnarray}
\end{subequations}
where the retarded form of $f(\omega)$ is given in Eq. (\ref{30})
and the non-retarded one in Eq. (\ref{32}).

Substituting $f(\omega)$ by $e^{-i\omega t} f(\omega)$ we can follow, 
in principle, the procedure outlined between Eqs. (\ref{36}) and (\ref{39}) 
of the previous section to extend the integration over real frequencies 
to an integration over the contour $C$ in Fig. 2.
The trouble is that on the lhs of (\ref{39}) we end up with the
integration over $\omega$ of $1/\omega$ multiplied by
\begin{equation} \label{56}
{\rm Im}\, e^{-i\omega t} f(\omega) =
\cos(\omega t) {\rm Im}\, f(\omega) + \sin(\omega t) {\rm Re}\, f(\omega) .
\end{equation}
Comparing this expression with the representation (\ref{55b}) we see that
only the first term is needed.
Within the $C$-contour formalism, we did not find a way how to get off 
the second (unwanted) term.

\begin{figure}
\begin{center}
\includegraphics[width=0.45\textwidth,clip]{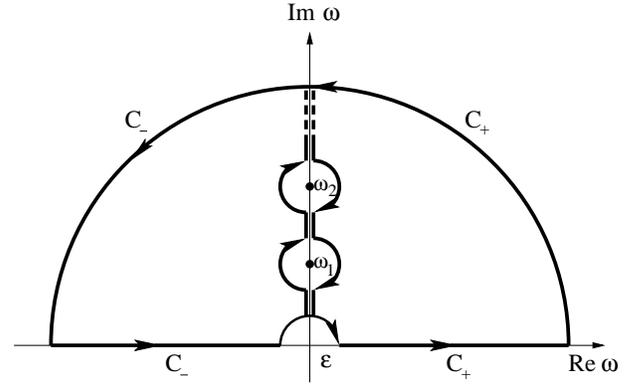}
\caption{Two contours in the complex frequency plane for $t\ne 0$;
$\omega_j=i\xi_j$ $(j=1,2,\ldots)$ are the Matsubara frequencies}
\end{center}
\end{figure}

Our new strategy is based on a transition from real times $t\ge 0$ 
to imaginary times.
In order to keep the function $e^{-i\omega t} f(\omega)/\omega$ integrable 
over real $\omega$, we make the substitutions $t\to -i\tau$ for $\omega>0$ and 
$t\to i\tau$ for $\omega<0$ $(\tau\ge 0)$, transforming in this way 
$e^{-i\omega t} f(\omega) \to e^{-\vert\omega\vert \tau} f(\omega)$.
In the $\omega$ upper half-plane, the contour $C$ in Fig. 2 will 
be replaced by two contours $C_+$ and $C_-$ in the quarter-spaces
${\rm Re}\,\omega>0$ and ${\rm Re}\,\omega<0$, respectively (see Fig. 3).
The contours are constructed in such a way that one avoids the singularities
(simple poles) of $f(\omega)/\omega$, the one at $\omega=0$ and the infinite 
sequence of poles at the imaginary Matsubara frequencies 
$\{ i\xi_j \}_{j=1}^{\infty}$.
The contour $C_+$ is directed along the real axis from 0 to $\infty$, 
except for an infinitesimal quarter-circle around the origin $\omega=0$,
continues by a quarter-circle at infinity and returns to the origin
neighborhood along the imaginary axis, avoiding by infinitesimal
semicircles the Matsubara frequencies $\{ i\xi_j \}$.
The contour $C_-$ is directed along the real axis from $-\infty$ to 0, 
except for an infinitesimal quarter-circle around the origin $\omega=0$,
continues along the imaginary axis, avoiding by infinitesimal
semicircles the Matsubara frequencies $\{ i\xi_j \}$, and returns to 
the starting point by a quarter-circle at infinity.
In close analogy with the procedure outlined between Eqs. (\ref{36}) 
and (\ref{39}), we can derive the integral equality
\begin{eqnarray}
& & \oint_{C_+} \frac{d\omega}{\omega} e^{-\omega \tau} f(\omega)
+ \oint_{C_-} \frac{d\omega}{\omega} e^{\omega \tau} f(\omega)
\nonumber \\ 
& = & - i \pi f(0) + {\cal P} \int_{-\infty}^{\infty} \frac{d\omega}{\omega} 
e^{-\vert \omega\vert \tau} f(\omega) \nonumber \\
& & - i \pi \sum_{j=1}^{\infty} \frac{{\rm Res}(f,i\xi_j)}{i\xi_j}
\left( e^{-i\xi_j \tau} + e^{i\xi_j \tau} \right) . \label{57}
\end{eqnarray}
Here, we have used two facts: the contributions coming from the straight-line 
fragments of the paths $C_+$ and $C_-$ between two neighboring frequencies 
$(i\xi_j,i\xi_{j+1})$ cancel exactly with one another due to the opposite 
direction of the integration and the contribution of the $C_+$-semicircle 
($C_-$-semicircle) around the pole $i\xi_j$ is equal to 
$-i\pi {\rm Res}(f,i\xi_j)/i\xi_j$ multiplied by the corresponding 
time-dependent factor $e^{-i\xi_j \tau}$ ($e^{i\xi_j \tau}$).
The lhs of Eq. (\ref{57}) is equal to zero since there are no poles
of the integrated functions inside the contours $C_+$ and $C_-$.
Since the ratio ${\rm Res}(f,i\xi_j)/i\xi_j$ is a real number,
taking the imaginary part of Eq. (\ref{57}) leads to
\begin{subequations}
\begin{equation} \label{58a}
\int_{-\infty}^{\infty} \frac{d\omega}{\omega} e^{-\vert \omega\vert \tau} 
{\rm Im}\, f(\omega) =  \pi f(0) + F(\tau,q) ,
\end{equation}
where
\begin{equation} \label{58b}
F(\tau,q) =  2 \pi \sum_{j=1}^{\infty} 
\frac{{\rm Res}(f,i\xi_j)}{i\xi_j} \cos(\xi_j \tau) .
\end{equation}
\end{subequations}

The function $F(\tau,q)$ represents the (imaginary) time-generalization 
of the static $F(0,q)$ (\ref{47b}).
In the retarded case, the generalization of the static formula (\ref{48}) 
reads
\begin{eqnarray}
F_{\rm qu}^{(\rm r)}(\tau,q) & = &  \frac{q^2}{2\pi} \sum_{j=1}^{\infty}
\frac{1}{\kappa_1(i\xi_j) \epsilon_2(i\xi_j) +
\kappa_2(i\xi_j) \epsilon_1(i\xi_j)} \nonumber \\ & & \times
\frac{[\epsilon_1(i\xi_j)-\epsilon_2(i\xi_j)]^2}{\epsilon_1(i\xi_j)
\epsilon_2(i\xi_j)} \cos(\xi_j \tau) . \label{59}
\end{eqnarray} 
In the non-retarded case, the generalization of the static formula (\ref{51}) 
reads
\begin{eqnarray}
F_{\rm qu}^{(\rm nr)}(t,q) & = & \frac{q}{2\pi} \sum_{j=1}^{\infty}
\left( \frac{1}{\epsilon_1(i\xi_j)} + \frac{1}{\epsilon_2(i\xi_j)}
\right. \nonumber \\ & & \left. 
- \frac{4}{\epsilon_1(i\xi_j)+\epsilon_2(i\xi_j)} \right) \cos(\xi_j \tau) . 
\label{60}
\end{eqnarray}
We have shown in Sec. IV that the series determining the functions 
in (\ref{59}) and (\ref{60}) are convergent for $\tau=0$.
The presence of the oscillating factor $\cos(\xi_j \tau)$ in the series 
for $\tau>0$ even improves their convergence property.
Let us ``pretend'' that we have found the explicit form of the function 
$F(\tau,q)$ (\ref{58b}), the type of the regime is irrelevant.  
To express the time-dependent correlation function (\ref{55b}), we have 
to return from imaginary to real times by considering the substitution 
$\tau\to it$ in Eq. (\ref{58a}).
Since the integral on the lhs of (\ref{58a}) is finite for all complex $\tau$
with ${\rm Re}\, \tau \ge 0$, there must exist a well-behaved analytic 
continuation $F(it,q)$ of the function $F(\tau,q)$.
Consequently,
\begin{eqnarray}
\beta S_{\rm qu}(t,q) & = & \frac{q}{4\pi} 
\left( \frac{1}{\epsilon_1(0)} + \frac{1}{\epsilon_2(0)} 
- \frac{4}{\epsilon_1(0)+\epsilon_2(0)} \right) \nonumber \\ & &
+ {\rm Re}\, F(it,q) . \label{61} 
\end{eqnarray}
For $t=0$, the operation ${\rm Re}$ on $F(0,q)$ become superfluous and we
recover the static result (\ref{47a}).

To check that the formalism works correctly, we reconsider 
the non-retarded regime for the contact between the jellium and vacuum.
Inserting the corresponding dielectric constants (\ref{52}) into (\ref{60})
and using the summation formula \cite{Gradshteyn}
\begin{equation} \label{62}
\sum_{j=1}^{\infty} \frac{(\alpha/\pi)^2}{j^2+(\alpha/\pi)^2}
\cos\left( j \frac{\pi}{\alpha} \omega \tau \right) = \frac{1}{2} \left[ 
\alpha \frac{\cosh(\alpha-\omega \tau)}{\sinh\alpha} - 1 \right]
\end{equation}
we obtain
\begin{eqnarray}
F_{\rm qu}^{(\rm nr)}(\tau,q) & = & \frac{q}{4\pi}
\left( \beta\hbar\omega_s \frac{
\cosh[(\beta\hbar\omega_s/2)-\omega_s \tau]}{\sinh(\beta\hbar\omega_s/2)}
\right. \nonumber \\ & & \left.
- \frac{\beta\hbar\omega_p}{2} \frac{
\cosh[(\beta\hbar\omega_p/2)-\omega_p \tau]}{\sinh(\beta\hbar\omega_p/2)}
- 1 \right) . \phantom{aaa} \label{63}
\end{eqnarray}
The analytic continuation of this function from $\tau$ to $i t$ is
well defined.
Using the relation ${\rm Re}\, \cosh(\alpha-i\omega t) = \cosh\alpha
\cos(\omega t)$ valid for real $\alpha$ and $\omega t$, we get 
\begin{eqnarray}
{\rm Re}\, F_{\rm qu}^{(\rm nr)}(it,q) & = & \frac{q}{4\pi}
[ 2 g(\omega_s) \cos(\omega_s t) \nonumber \\
& & - g(\omega_p) \cos(\omega_p t) -1] \label{64} 
\end{eqnarray}
with $g(\omega)$ defined in (\ref{8b}).
Adding to this result the first term $q/(4\pi)$ in (\ref{61}),
we reproduce the previous time-dependent result (\ref{8a}). 

In the retarded case, the function $F_{\rm qu}^{(\rm r)}(\tau,q)$ in
(\ref{59}) is of the order $O(q^2)$.
The same property holds for its analytic continuation 
$F_{\rm qu}^{(\rm r)}(i t,q)$.
Eq. (\ref{61}) then tells us that, in the limit $q\to 0$,
\begin{equation} \label{65}
\beta S_{\rm qu}^{(\rm r)}(t,q) = \frac{q}{4\pi} 
\left( \frac{1}{\epsilon_1(0)} + \frac{1}{\epsilon_2(0)} 
- \frac{4}{\epsilon_1(0)+\epsilon_2(0)} \right) .
\end{equation}
In view of the representation (\ref{6}), the quantum time-dependent prefactor 
to the asymptotic decay takes, for any temperature, its static classical form
\begin{equation} \label{66}
h_{\rm qu}^{(\rm r)}(t) = - \frac{1}{8\pi^2} 
\left( \frac{1}{\epsilon_1(0)} + \frac{1}{\epsilon_2(0)} 
- \frac{4}{\epsilon_1(0)+\epsilon_2(0)} \right) .
\end{equation}
This result holds for all possible media combinations if one takes
$\epsilon(0)\to i\infty$ for conductors, $\epsilon(0) = \epsilon_0>1$ 
for dielectrics and $\epsilon(0) = 1$ for vacuum.

\section{Conclusion}
Although the present work is rather technical, its main result (\ref{65}) 
(or, equivalently, (\ref{66})) is of physical interest.
It represents the generalization of the analogous result, obtained in 
Refs. \cite{SJ,JS} and valid exclusively for the special jellium model of 
conductors in contact with vacuum, to all possible media 
(conductor, dielectric, vacuum) configurations.
The derivation of the general result (\ref{65}), based on the integration 
in the complex plane and on the known analytic properties of dielectric 
functions in the frequency upper half-plane, is much simpler and more
transparent than the one for the special case of the jellium in contact 
with vacuum \cite{SJ,JS}.

There is still an open question: which are {\em physical} reasons that 
the inclusion of retardation effects causes the asymptotic decay of 
time-dependent quantum surface charge correlations to take its
static classical form, independent of $\hbar$ and $c$? 
For the static quantum case $t=0$, a possible explanation might be that the 
surface charge correlation function depends on the only dimensionless
quantity $\beta\hbar c/R$ constructed from universal constant and the distance.
If this is the case, the large-distance asymptotics $R\to\infty$
is equivalent to the classical limit $\beta\hbar\to 0$.
A verification whether this claim is true or not is left for the future. 
The case $t\ne 0$ will be still an open problem.
It might be also possible to verify our results by using
a fully microscopic approach.

We believe that Rytov's fluctuational electrodynamics and the techniques
presented are applicable also to other important phenomena associated
with the presence of a surface between distinct media.
One of such problems can be the evaluation of the shape-dependent 
dielectric susceptibility tensor \cite{LL,Choquard,Jancovici04} 
for quantum Coulomb fluids, with and without retardation effects.
Another interesting topic is the large-distance behavior of 
the current-current correlation function near an interface between two media. 

\begin{acknowledgments}
L. \v{S}. is grateful to LPT for very kind invitation and hospitality.
The support received from the European Science Foundation (ESF
``Methods of Integrable Systems, Geometry, Applied Mathematics''),
Grant VEGA No. 2/0113/2009 and CE-SAS QUTE is acknowledged. 
\end{acknowledgments}

\bibliography{surface3}

\begin{thebibliography}{19}
\expandafter\ifx\csname natexlab\endcsname\relax\def\natexlab#1{#1}\fi
\expandafter\ifx\csname bibnamefont\endcsname\relax
  \def\bibnamefont#1{#1}\fi
\expandafter\ifx\csname bibfnamefont\endcsname\relax
  \def\bibfnamefont#1{#1}\fi
\expandafter\ifx\csname citenamefont\endcsname\relax
  \def\citenamefont#1{#1}\fi
\expandafter\ifx\csname url\endcsname\relax
  \def\url#1{\texttt{#1}}\fi
\expandafter\ifx\csname urlprefix\endcsname\relax\def\urlprefix{URL }\fi
\providecommand{\bibinfo}[2]{#2}
\providecommand{\eprint}[2][]{\url{#2}}

\bibitem[{\citenamefont{\v{S}amaj and Jancovici}(2008)}]{SJ}
\bibinfo{author}{\bibfnamefont{L.}~\bibnamefont{\v{S}amaj}} \bibnamefont{and}
  \bibinfo{author}{\bibfnamefont{B.}~\bibnamefont{Jancovici}},
  \bibinfo{journal}{Phys. Rev. E} \textbf{\bibinfo{volume}{78}},
  \bibinfo{pages}{051119} (\bibinfo{year}{2008}).

\bibitem[{\citenamefont{Jancovici and \v{S}amaj}(2009)}]{JS}
\bibinfo{author}{\bibfnamefont{B.}~\bibnamefont{Jancovici}} \bibnamefont{and}
  \bibinfo{author}{\bibfnamefont{L.}~\bibnamefont{\v{S}amaj}},
  \bibinfo{journal}{Phys. Rev. E} \textbf{\bibinfo{volume}{79}},
  \bibinfo{pages}{021111} (\bibinfo{year}{2009}).

\bibitem[{\citenamefont{Jackson}(1975)}]{Jackson}
\bibinfo{author}{\bibfnamefont{J.~D.} \bibnamefont{Jackson}},
  \emph{\bibinfo{title}{Classical Electrodynamics}}
  (\bibinfo{publisher}{Wiley}, \bibinfo{address}{New York},
  \bibinfo{year}{1975}), \bibinfo{edition}{2nd} ed.

\bibitem[{\citenamefont{Jancovici}(1982)}]{Jancovici82}
\bibinfo{author}{\bibfnamefont{B.}~\bibnamefont{Jancovici}},
  \bibinfo{journal}{J. Stat. Phys.} \textbf{\bibinfo{volume}{29}},
  \bibinfo{pages}{263} (\bibinfo{year}{1982}).

\bibitem[{\citenamefont{Jancovici}(1995)}]{Jancovici95}
\bibinfo{author}{\bibfnamefont{B.}~\bibnamefont{Jancovici}},
  \bibinfo{journal}{J. Stat. Phys.} \textbf{\bibinfo{volume}{80}},
  \bibinfo{pages}{445} (\bibinfo{year}{1995}).

\bibitem[{\citenamefont{Jancovici}(1985)}]{Jancovici85a}
\bibinfo{author}{\bibfnamefont{B.}~\bibnamefont{Jancovici}},
  \bibinfo{journal}{J. Stat. Phys.} \textbf{\bibinfo{volume}{39}},
  \bibinfo{pages}{427} (\bibinfo{year}{1985}).

\bibitem[{\citenamefont{Jancovici et~al.}(1985)\citenamefont{Jancovici,
  Lebowitz, and Martin}}]{Jancovici85b}
\bibinfo{author}{\bibfnamefont{B.}~\bibnamefont{Jancovici}},
  \bibinfo{author}{\bibfnamefont{J.~L.} \bibnamefont{Lebowitz}},
  \bibnamefont{and} \bibinfo{author}{\bibfnamefont{P.~A.}
  \bibnamefont{Martin}}, \bibinfo{journal}{J. Stat. Phys.}
  \textbf{\bibinfo{volume}{41}}, \bibinfo{pages}{941} (\bibinfo{year}{1985}).

\bibitem[{\citenamefont{Jancovici et~al.}(1995)\citenamefont{Jancovici,
  Lebowitz, and Martin}}]{Jancovicicor}
\bibinfo{author}{\bibfnamefont{B.}~\bibnamefont{Jancovici}},
  \bibinfo{author}{\bibfnamefont{J.~L.} \bibnamefont{Lebowitz}},
  \bibnamefont{and} \bibinfo{author}{\bibfnamefont{P.~A.}
  \bibnamefont{Martin}}, \bibinfo{journal}{J. Stat. Phys.}
  \textbf{\bibinfo{volume}{79}}, \bibinfo{pages}{789(E)}
  (\bibinfo{year}{1995}).

\bibitem[{\citenamefont{Landau and Lifshitz}(1960)}]{LL}
\bibinfo{author}{\bibfnamefont{L.}~\bibnamefont{Landau}} \bibnamefont{and}
  \bibinfo{author}{\bibfnamefont{E.}~\bibnamefont{Lifshitz}},
  \emph{\bibinfo{title}{Electrodynamics of Continuous Media}}
  (\bibinfo{publisher}{Pergamon Press}, \bibinfo{address}{Oxford},
  \bibinfo{year}{1960}).

\bibitem[{\citenamefont{Rytov}(1958)}]{Rytov}
\bibinfo{author}{\bibfnamefont{S.}~\bibnamefont{Rytov}}, \bibinfo{journal}{Sov.
  Phys. JETP} \textbf{\bibinfo{volume}{6}}, \bibinfo{pages}{130}
  (\bibinfo{year}{1958}).

\bibitem[{\citenamefont{Alastuey}()}]{Alastuey08}
\bibinfo{author}{\bibfnamefont{A.}~\bibnamefont{Alastuey}}, \eprint{private
  communication}.

\bibitem[{\citenamefont{Bohr}()}]{Bohr}
\bibinfo{author}{\bibfnamefont{N.}~\bibnamefont{Bohr}}, \eprint{{\it
  Dissertation} (Copenhagen, unpublished) 1911}.

\bibitem[{\citenamefont{Leeuwen}(1921)}]{Leeuwen}
\bibinfo{author}{\bibfnamefont{J.~H.~V.} \bibnamefont{Leeuwen}},
  \bibinfo{journal}{J. Physique} \textbf{\bibinfo{volume}{2}},
  \bibinfo{pages}{361} (\bibinfo{year}{1921}).

\bibitem[{\citenamefont{Alastuey and Appel}(2000)}]{Alastuey00}
\bibinfo{author}{\bibfnamefont{A.}~\bibnamefont{Alastuey}} \bibnamefont{and}
  \bibinfo{author}{\bibfnamefont{W.}~\bibnamefont{Appel}},
  \bibinfo{journal}{Physica A} \textbf{\bibinfo{volume}{276}},
  \bibinfo{pages}{508} (\bibinfo{year}{2000}).

\bibitem[{\citenamefont{Lifshitz and Pitaevskii}(1980)}]{LP}
\bibinfo{author}{\bibfnamefont{E.~M.} \bibnamefont{Lifshitz}} \bibnamefont{and}
  \bibinfo{author}{\bibfnamefont{L.~P.} \bibnamefont{Pitaevskii}},
  \emph{\bibinfo{title}{Statistical Physics, Part 2, Chapter VIII}}
  (\bibinfo{publisher}{Pergamon Press}, \bibinfo{address}{Oxford},
  \bibinfo{year}{1980}).

\bibitem[{\citenamefont{Gradshteyn and Ryzhik}(2000)}]{Gradshteyn}
\bibinfo{author}{\bibfnamefont{I.~S.} \bibnamefont{Gradshteyn}}
  \bibnamefont{and} \bibinfo{author}{\bibfnamefont{I.~M.}
  \bibnamefont{Ryzhik}}, \emph{\bibinfo{title}{Table of Integrals, Series, and
  Products}} (\bibinfo{publisher}{Academic Press}, \bibinfo{address}{London},
  \bibinfo{year}{2000}), \bibinfo{edition}{6th} ed.

\bibitem[{\citenamefont{Pitarke et~al.}(2007)\citenamefont{Pitarke, Silkin,
  Chulkov, and Echenique}}]{Pitarke}
\bibinfo{author}{\bibfnamefont{J.~M.} \bibnamefont{Pitarke}},
  \bibinfo{author}{\bibfnamefont{V.~M.} \bibnamefont{Silkin}},
  \bibinfo{author}{\bibfnamefont{E.~V.} \bibnamefont{Chulkov}},
  \bibnamefont{and} \bibinfo{author}{\bibfnamefont{P.~M.}
  \bibnamefont{Echenique}}, \bibinfo{journal}{Rep. Prog. Phys.}
  \textbf{\bibinfo{volume}{70}}, \bibinfo{pages}{1} (\bibinfo{year}{2007}).

\bibitem[{\citenamefont{Choquard et~al.}(1989)\citenamefont{Choquard, Piller,
  Rentsch, and Vieillefosse}}]{Choquard}
\bibinfo{author}{\bibfnamefont{P.}~\bibnamefont{Choquard}},
  \bibinfo{author}{\bibfnamefont{B.}~\bibnamefont{Piller}},
  \bibinfo{author}{\bibfnamefont{R.}~\bibnamefont{Rentsch}}, \bibnamefont{and}
  \bibinfo{author}{\bibfnamefont{P.}~\bibnamefont{Vieillefosse}},
  \bibinfo{journal}{J. Stat. Phys.} \textbf{\bibinfo{volume}{55}},
  \bibinfo{pages}{1185} (\bibinfo{year}{1989}).

\bibitem[{\citenamefont{Jancovici and \v{S}amaj}(2004)}]{Jancovici04}
\bibinfo{author}{\bibfnamefont{B.}~\bibnamefont{Jancovici}} \bibnamefont{and}
  \bibinfo{author}{\bibfnamefont{L.}~\bibnamefont{\v{S}amaj}},
  \bibinfo{journal}{J. Stat. Phys.} \textbf{\bibinfo{volume}{114}},
  \bibinfo{pages}{1211} (\bibinfo{year}{2004}).

\end{thebibliography}

\end{document}